\documentclass[aps,prl,longbibliography,twocolumn,superscriptaddress]{revtex4-2}
\usepackage{amsmath,amssymb,bm,graphicx,color,bbold,hyperref,keyval,url,latexsym,dsfont}
\usepackage{xcolor}
\usepackage[normalem]{ulem} 
\usepackage{CJK}
\usepackage{float}
\usepackage{comment}
\usepackage[version=4]{mhchem}

\begin{document}
\title{Quantifying the Phase Diagram and Hamiltonian of $S=1/2$ Kagome Antiferromagnets: Bridging Theory and Experiment}

\begin{CJK*}{UTF8}{}
\author{Shengtao Jiang (\CJKfamily{gbsn}蒋晟韬)}
\affiliation{Stanford Institute for Materials and Energy Sciences, SLAC National Accelerator Laboratory, Menlo Park, CA 94025, USA}

\author{Arthur C. Campello}

\affiliation{Stanford Institute for Materials and Energy Sciences, SLAC National Accelerator Laboratory, Menlo Park, CA 94025, USA}
\affiliation{Department of Applied Physics, Stanford University, Stanford, CA 94305, USA}

\author{Wei He}
\affiliation{Stanford Institute for Materials and Energy Sciences, SLAC National Accelerator Laboratory, Menlo Park, CA 94025, USA}

\author{Jiajia Wen}
\affiliation{Stanford Institute for Materials and Energy Sciences, SLAC National Accelerator Laboratory, Menlo Park, CA 94025, USA}

\author{Daniel M. Pajerowski}
\affiliation{Neutron Scattering Division, Oak Ridge National Laboratory, Oak Ridge TN 37830, USA}

\author{Young S. Lee}
\affiliation{Stanford Institute for Materials and Energy Sciences, SLAC National Accelerator Laboratory, Menlo Park, CA 94025, USA}
\affiliation{Department of Applied Physics, Stanford University, Stanford, CA 94305, USA}

\author{Hong-Chen Jiang}
\altaffiliation{hcjiang@stanford.edu}
\affiliation{Stanford Institute for Materials and Energy Sciences, SLAC National Accelerator Laboratory, Menlo Park, CA 94025, USA}
\date{\today}

\begin{abstract}
Spin-$1/2$ kagome antiferromagnets are leading candidates for realizing quantum spin liquid (QSL) ground states. While QSL ground states are predicted for the pure Heisenberg model, understanding the robustness of the QSL to additional interactions that may be present in real materials is a forefront question in the field. Here we employ large-scale density-matrix renormalization group simulations to investigate the effects of next-nearest neighbor exchange couplings $J_2$ and Dzyaloshinskii-Moriya interactions $D$, which are relevant to understanding the prototypical kagome materials herbertsmithite and Zn-barlowite. By utilizing clusters as large as XC12 and extrapolating the results to the thermodynamic limit, we precisely delineate the scope of the QSL phase, which remains robust across an expanded parameter range of $J_2$ and $D$. Direct comparison of the simulated static and dynamic spin structure factors with inelastic neutron scattering reveals the parameter space of the Hamiltonians for herbertsmithite and Zn-barlowite, and, importantly, provides compelling evidence that both materials exist within the QSL phase. These results establish a powerful convergence of theory and experiment in this most elusive state of matter.
\end{abstract}
\maketitle
\end{CJK*}

{\it Introduction.} Quantum spin liquids (QSLs) are exotic phases of matter that avoid spontaneous symmetry breaking even at zero temperature and support fractionalized excitations~\cite{Anderson1973,Kivelson1987,Rokhsar1988,Laughlin1988,Kivelson1989,Moessner2001,Balents2010,Savary2016,Broholm2019}.
A prominent system known to host a QSL is the kagome lattice antiferromagnet, characterized by strong quantum fluctuations induced by pronounced geometric frustration.
Indeed, a consensus has been reached that the kagome antiferromagnetic (AF) Heisenberg model with nearest-neighbor (NN) exchange coupling $J_1$ has a QSL ground state, although 
its precise nature, i.e., gapped (such as a $Z_2$ QSL) or gapless (such as a $U(1)$ Dirac QSL), remains under debate~\cite{Jiang2008,Yan2011,Jiang2012,Depenbrock2012,Mei2017,Ran2007,Liao2017,He2017,Gong2015,Haghshenas2019,Zhu2019,Lauchli2011,Lauchli2019,sun2024possible,ed1,vmctaoli}.

On the experimental side, significant progress has been made in the synthesis and investigation of QSL candidate materials. A leading example is herbertsmithite ZnCu$_3$(OH)$_6$Cl$_2$~\cite{Helton2007,Mendels2007,Han2012,Norman2016}, in which the spin-1/2 moments on Cu$^{2+}$ are arranged on a structurally perfect kagome lattice, with nonmagnetic Zn$^{2+}$ ions separating those kagome planes. Experiments indicate a dominant NN AF coupling $J_1\approx 17$~meV in herbertsmithite~\cite{Helton2007, Han2012}, which establishes it as an ideal platform for realizing and exploring the QSL state. 
This is indeed evidenced by the experimental observations demonstrating the absence of magnetic order down to temperatures as low as $50$~mK~\cite{Helton2007,Mendels2007,Vries2009}.
Zn-barlowite (\ce{ZnCu3(OH)6FBr}) is another well-established kagome QSL candidate material~\cite{Feng2017, Smaha2018, Smaha2020, Tustain2020Magnetic, Fu2021Dynamic} with a dominant $J_1\approx 14$~meV~\cite{Smaha2020}.
For both materials, fractionalized spin excitations have been observed in inelastic neutron scattering (INS) measurements on single crystal samples consistent with a QSL state~\cite{Han2012,Han2016,Breidenbach2025,HBS_unpublished,gapless-exp1}. Furthermore, experiments suggest a finite spin gap for the intrinsic kagome moments in both materials~\cite{Han2016,Fu2015,Wang2021,Breidenbach2025}. 

Despite significant recent theoretical and experimental progress, further steps are necessary to bridge the gap between the QSL in the $J_1$-only kagome AF model and real materials where additional interactions may also be present. Indeed, it has been proposed that a minimal spin model for the two kagome materials 
include a small but finite next-nearest-neighbor (NNN) Heisenberg interaction $J_2$~\cite{Han2012,Jeschke2013}. Meanwhile, the bulk magnetic properties, anisotropies in thermodynamic quantities, and electron spin resonance measurements  
suggest the presence of a weak Dzyaloshinskii-Moriya (DM) interaction~\cite{Zorko2008,Han2012a, Rigol2007,Messio2010}, which arises from spin-orbit coupling~\cite{Lee2013}. 
The effects of the $J_2$ or DM interaction have been explored in various contexts using semi-classical or numerical approaches~\cite{harris1992,chubukov1992,Elhajal2002,huh2010,Cepas2008,sasha2015,Gong2015,Kolley2015,vmc1,vmc2}.
In particular, it is crucial to investigate how these additional interactions influence the stability of the QSL state and establish the microscopic minimal model for herbertsmithite and Zn-barlowite on a quantitative level.

\begin{figure}[t]
  \includegraphics[width= 1.0\linewidth]{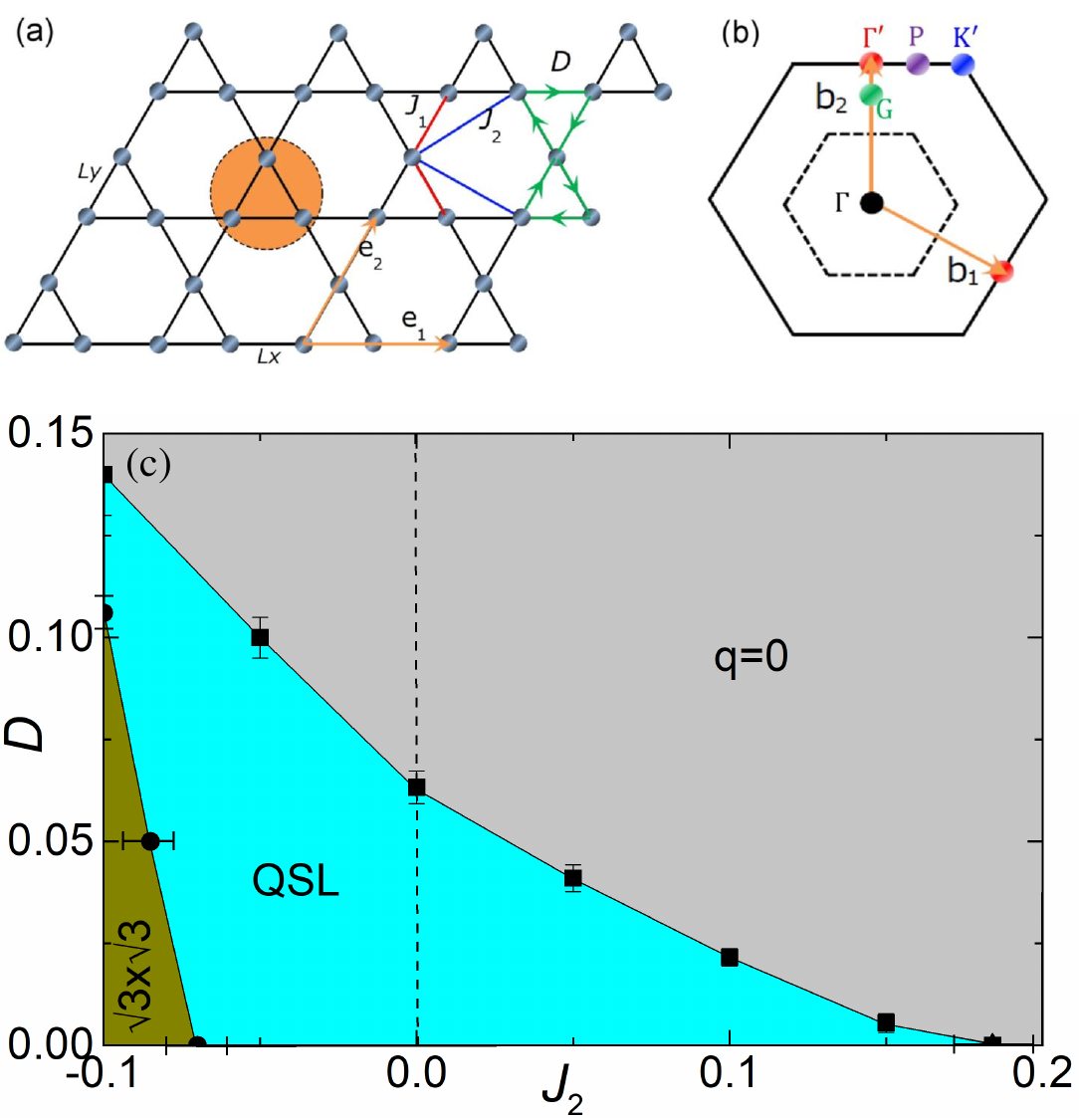}
  \vskip -0.2cm
  \caption{(a) Heisenberg $J_1$-$J_2$-$D$ model in Eq.(\ref{Eq:Ham}) on a kagome cylinder. Periodic and open boundary conditions are imposed, respectively, along the directions specified by the lattice basis vectors $e_2$ and $e_1$.
  The small triangle in the shaded region denotes a unit cell.
  $J_1$ and $J_2$ are NN and NNN spin exchange couplings, and $D$ is the DM interaction. (b) The first and extended Brillouin zones, the reciprocal lattice vectors $b_1$ and $b_2$, and the high-symmetry points: $\Gamma$, $\Gamma'$, $\rm K^\prime$, P (the midpoint of $\Gamma'$, $\rm K^\prime$), and $G=5b_2/6$.
  (c) Ground state phase diagram of the system as a function of $J_2$ and $D$. The solid symbols with error bars label the phase boundaries determined by DMRG calculations.}
  \label{Fig:latt_and_phd}
\end{figure}

In this Letter, we address these questions and bridge the gap between our understanding of QSL's in models and real materials using large-scale density-matrix renormalization group (DMRG)~\cite{White1992,White1993} simulations with input from neutron scattering measurements. We establish the ground-state phase diagram of the kagome AFM model in the $J_2$ and $D$ plane, demonstrating that the QSL phase remains robust in the presence of both interactions. Specifically, the QSL phase is stable  within the ranges $-0.07 \lesssim J_2 \lesssim 0.18$ and $0 \lesssim D \lesssim 0.06$ 
as shown in Fig.~\ref{Fig:latt_and_phd}(c).
By comparing both the static and dynamic spin structural factors with neutron scattering results, we find that the model parameters for both materials are constrained to similar regions, and both fall entirely within the QSL phase.

\textit{Model.}--
The kagome lattice Heisenberg $J_1$-$J_2$ model with a DM interaction, depicted in Fig.~\ref{Fig:latt_and_phd}(a), is defined as:
\begin{equation}
    \label{Eq:Ham}
H = J_1\sum_{\langle ij\rangle} \mathbf{S}_i\cdot \mathbf{S}_j + J_2\sum_{\langle \langle ij\rangle \rangle} \mathbf{S}_i\cdot \mathbf{S}_j
+ \sum_{\langle ij\rangle}\mathbf{D}_{ij}\cdot (\mathbf{S}_i\times \mathbf{S}_j ).
\end{equation}
Here $\mathbf{S}_i$ is the S=1/2 spin operator on site $i$, and the first two terms denote spin exchange couplings between NN and NNN sites, respectively. The third term is the DM interaction originating from spin-orbit coupling, which can be present when lattice inversion symmetry is broken~\cite{Dzyaloshinsky1958,Moriya1960,Lee2013}. The DM vector $\mathbf{D}_{ij}$ depends on the convention of the bond orientation as $\mathbf{D}_{ij}=-\mathbf{D}_{ji}$~\cite{Elhajal2002,Cepas2008,Zorko2008}. For a given convention~\cite{Lee2013}, the DM vectors are shown in Fig.\ref{Fig:latt_and_phd}(a).
Since electron-spin resonance measurements suggest the existence of a leading out-of-plane component of $\mathbf{D}_{ij}$~\cite{Zorko2008}, we hence choose $\mathbf{D}_{ij}=D\hat{z}$ in the present study. 
This reduces the computational costs and facilitates reliable DMRG simulations on large systems by leveraging the $U(1)$ spin rotational symmetry. We adopt the convention where $D>0$ corresponds to all bonds $i\rightarrow j$ being oriented clockwise, as 
illustrated in Fig.\ref{Fig:latt_and_phd}(a)~\cite{Lee2013}. We set $J_1=1$ as an energy unit.

In the absence of $J_2$ and DM interactions, numerous numerical studies have consistently identified a QSL ground state~\cite{Jiang2008,Yan2011,Jiang2012,Depenbrock2012,Mei2017,Ran2007,Liao2017,He2017,sun2024possible,vmc1,vmc2,ed1,Gong2015}, although its precise nature, i.e., gapped or gapless in the thermodynamic limit, remains currently under debate. We do not aim to further address the gapped versus gapless nature of the QSL, as this question is beyond the scope of the current study. We are interested in examining the robustness of the QSL state, where previous DMRG studies suggest that the QSL remains the ground state in an extended parameter region between $J_2\leq 0.15\sim 0.20$ and $J_2\geq -0.1\sim -0.05$ in the absence of DM interaction~\cite{Gong2015,Kolley2015}.
However, the ground state phase diagram in the presence of {\it both} $J_2$ and DM interactions remains largely unexplored. 
Moreover, the effects of these parameters on the spin excitations of herbertsmithite and Zn-barlowite have yet to be quantitatively explored. This study aims to address these questions by simulating the spin Hamiltonian in Eq.(\ref{Eq:Ham}) and comparing the results with results from inelastic neutron scattering.

\begin{figure}[t]
  \includegraphics[width=\linewidth]{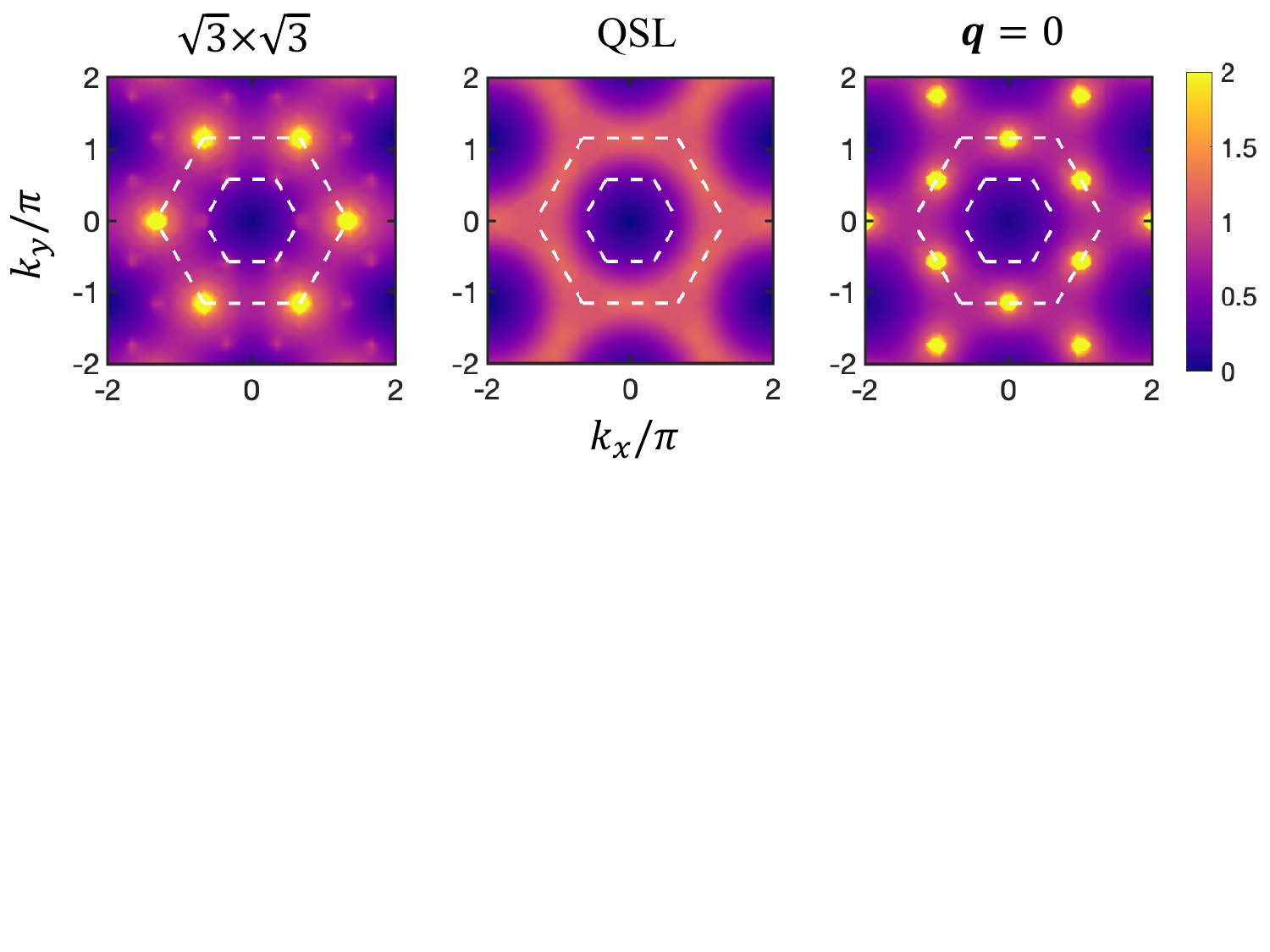}
  \vskip -3.5cm
  \caption{Static spin structure factors $S(\mathbf{q})$ obtained from ground state DMRG simulations on XC12-9 cylinders for three characteristic sets of parameters:
  (a) $J_2=-0.1$ and $D=0$, the system exhibits the $\sqrt{3}\times\sqrt{3}$ magnetic order with sharp peaks in $S(\mathbf{q})$ at $\rm K'$; (b)$J_2=0$ and $D=0$, the system has a QSL ground state with diffuse $S(\mathbf{q})$;
  (c) $J_2=0.1$ and $D=0.1$, the system has $q=0$ order with sharp peaks in $S(\mathbf{q})$ at $\rm \Gamma'$; The first and extended Brillouin zones are indicated by the dashed hexagons. The results have been $D_6$ symmetrized. The color scale has an upper cutoff of 2.
  }\label{Fig:Strfac}
\end{figure}

\textit{Phase diagram.}--
Based on large-scale DMRG simulations, we establish the ground state phase diagram of the model Hamiltonian in Eq.(\ref{Eq:Ham}) as shown in Fig.~\ref{Fig:latt_and_phd}(c). The phases are determined by calculating the equal-time spin-spin correlation function and the corresponding spin structure factors.
Our results reveal an extended QSL region that is characterized by diffuse structure factors without sharp peak-like features. Specifically, the QSL phase is stable within the range $-0.07(1)\lesssim J_2\lesssim 0.18(1)$ in the absence of DM interaction, and $D\lesssim 0.063(4)$ when $J_2$ is absent. 
These results align well with previous studies along the $J_2$ and $D$ axis, respectively~\cite{Kolley2015,Cepas2008}.
Outside the spin liquid region, we find that the system develops long-range magnetic order of the $\sqrt{3}\times \sqrt{3}$ type when $J_2$ is more ferromagnetic and $D$ is small. Conversely, for stronger AF $J_2$ and/or larger $D$, the system transitions into the $q=0$ ordered state. The characteristic spin structure factors for both the magnetically ordered phases and QSL phase are presented in Fig.\ref{Fig:Strfac}.
 
\textit{DMRG setup.}--
The lattice geometry used in our DMRG simulations is depicted in Fig.\ref{Fig:latt_and_phd}(a), where $\mathbf{e}_1=(2,0)$ and $\mathbf{e}_2=(1,\sqrt{3})$ denote the two basis vectors. We consider kagome cylinders with open and periodic boundary conditions along the $\mathbf{e}_1$ and $\mathbf{e}_2$ directions. Following the convention in Ref.~\cite{Yan2011,Kolley2015}, we refer to a cylinder with $L_x$ and $L_y$ unit cells ($2L_y$ and $2L_x$ sites) in the $\mathbf{e}_2$ and $\mathbf{e}_1$ directions as XC$2L_y$-$L_x$.
For a better connection to the two-dimensional (2D) limit~\cite{steve-sasha}, we primarily focus on ``square-like'' cylinders with aspect ratio $1< L_x/L_y\leq 2$ and width ranging from XC6 to XC12.
We perform up to 50 DMRG sweeps and keep up to $m=8000$ number of states in each DMRG block for the case of $D=0$ and $m=5120$ number of states for the case of $D>0$. These yield a typical truncation error $\epsilon\lesssim 10^{-5}$ with excellent convergence for our results when extrapolated to the limit $\epsilon=0$, i.e., $m=\infty$.

\textit{Phase determination using static structure factor.}-- We begin with measurements of the equal-time spin-spin correlation $\langle \mathbf{S}_i\cdot\mathbf{S}_j\rangle$ between sites $i$ and $j$, and the corresponding static spin structure factor $S(\mathbf{q})$, defined as
\begin{eqnarray}\label{Eq:Strfac}
S(\mathbf{q})=\frac{1}{N}\sum_{i,j=1}^{N}\langle \mathbf{S}_i\cdot\mathbf{S}_j\rangle e^{i\mathbf{q}\cdot(\mathbf{r}_i-\mathbf{r}_j)}.
\end{eqnarray}
Figure~\ref{Fig:Strfac} presents examples of $S(\mathbf{q})$ for the $\sqrt{3}\times\sqrt{3}$, QSL and $\mathbf{q}$=0 states, corresponding to characteristic parameter sets ($J_2=-0.1$, $D=0$), ($J_2=0$, $D=0$), and ($J_2=0.1$, $D=0.1$) on the XC12-9 cylinder.  The structure factor $S(\mathbf{q})$ has a clear peak at $\rm K'$ momenta in the extended Brillouin zone (BZ) for ($J_2=-0.1$, $D=0$), corresponding to the long-ranged $\sqrt{3}\times \sqrt{3}$ magnetic order. 
On the other hand, for ($J_2=0.1$, $D=0.1$), the structure factor is sharply peaked at $\Gamma^\prime$ momenta (the center of the second Brillouin zone). This aligns well with the $q=0$ magnetic ordered state, which preserves the translational symmetry of the system. In stark contrast, $S(\mathbf{q})$ becomes diffuse without any sharp peak for ($J_2=0$, $D=0$), demonstrating the absence of any long-range magnetic order, consistent with a QSL state. 

\begin{figure}
  \includegraphics[width=\linewidth]{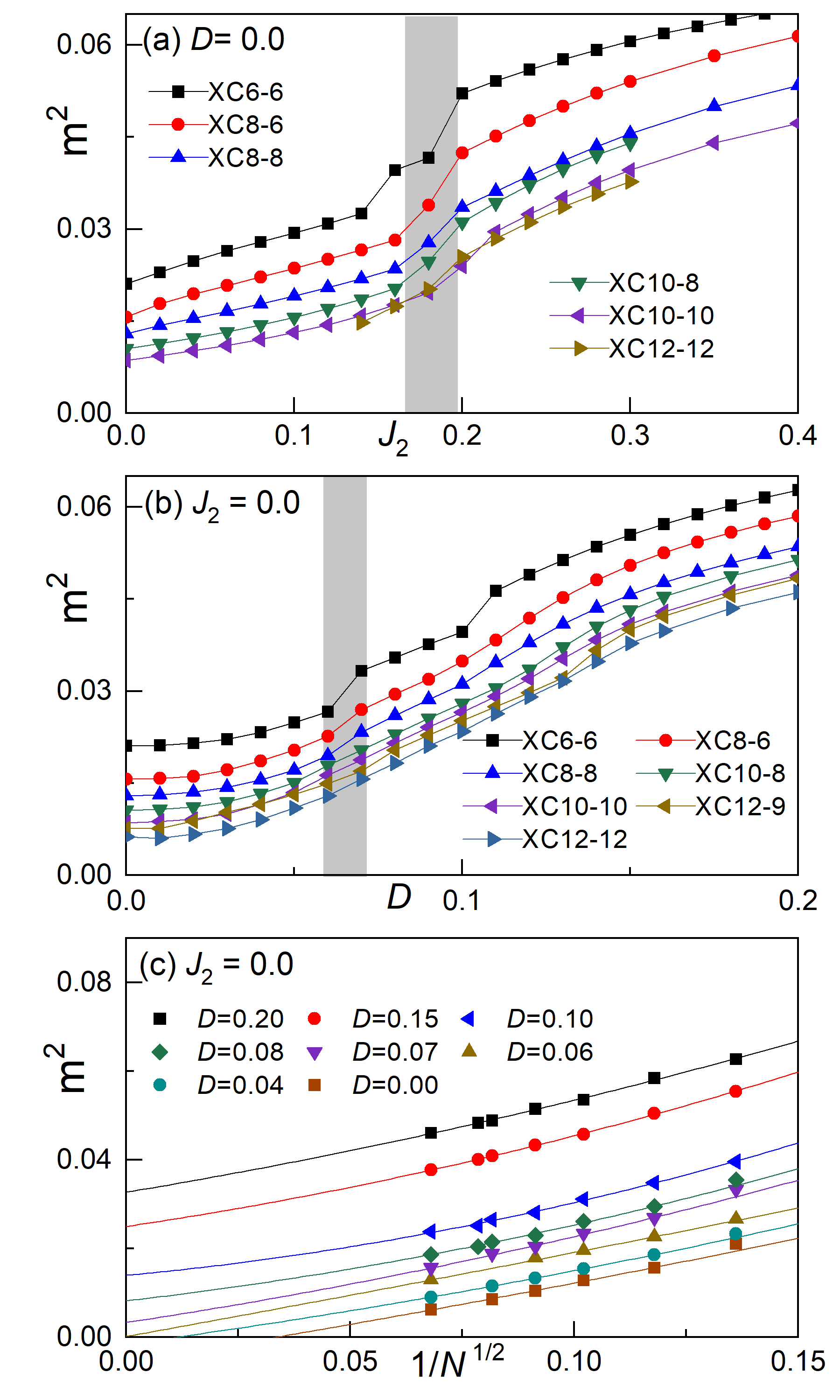}
  \caption{Squared $\mathbf{q}$=0 magnetic order parameter $m^2(\mathbf{Q}=\Gamma')$ as a function of (a) $J_2$ at $D=0$ and (b) $D$ at $J_2=0$. The shaded regions label the phase transitions between the QSL and $\mathbf{q}=0$ magnetically ordered phase. (c) Examples of finite-size extrapolations of $m^2(\mathbf{Q}=\Gamma')$ for different $D$ at $J_2=0$ using second-order polynomials in $1/\sqrt{N}$. 
}\label{Fig:Ms2FSS}
\end{figure}

We use two different ways to quantify the magnetic order and determine the phase boundaries. We first define the squared order parameter as $m^2(\mathbf{Q},N)=S(\mathbf{Q})/N$, where $\mathbf{Q}$ represents the peak positions associated with different magnetic orders, i.e., $\mathbf{Q}={\rm K'}$ for $\sqrt{3}\times\sqrt{3}$ magnetic order and $\mathbf{Q}={\Gamma '}$ for $q=0$ magnetic order. Figure~\ref{Fig:Ms2FSS}(a) and (b) show $m^2({\mathbf{Q}=\Gamma'})$ for the $q=0$ order as a function of $J_2$ at $D=0$ and as a function of $D$ at $J_2=0$ for different cylinders.
To quantitatively analyze the order, we perform an extrapolation of $m^2(\mathbf{Q},N)$ to the 2D limit 
as illustrated in Figure~\ref{Fig:Ms2FSS}(c), using results from various kagome cylinders ranging from XC6 to XC12. 
The extrapolated $m^2(\mathbf{Q})$ denotes the squared order parameter in the thermodynamic limit $N=\infty$, which remains finite in the magnetic ordered states but vanishes in the QSL state. 
Examples of finite-size extrapolations are shown in Fig. \ref{Fig:Ms2FSS}(c) for various $D$ at $J_2=0$.
It is evident that $m^2(\mathbf{Q}=\Gamma')$ remains finite for $D\gtrsim 0.07$, indicating the presence of the $q=0$ magnetic order. In contrast, for $D\leq 0.06$, $m^2(\mathbf{Q}=\Gamma')$ extrapolates to zero or even a negative value, indicating the absence of such an order. Consequently, the phase boundary between the $q=0$ and QSL phases is established at $D\approx 0.06$ along the $J_2$=0 line.

Alternatively, the phase boundary can be identified by examining $m^2(\mathbf{Q},N)$ and its derivative as functions of $J_2$ or $D$. For a given system size, $m^2(\mathbf{Q},N)$ remains relatively small in the QSL phase but becomes significantly larger in the magnetically ordered phases. Consequently, a sharp increase (or even a discontinuity in small systems) accompanied by a pronounced peak in its derivative is expected at the phase boundary as indicated by the shaded regions in Fig.~\ref{Fig:Ms2FSS}(a) and (b).
Using this approach, we determine the phase boundaries between the QSL and the $q=0$ ordered phase to be around $D = 0.065(5)$ in the absence of $J_2$, and $J_2 = 0.18(2)$ for $D=0$. Similarly, the boundary between the QSL and the $\sqrt{3} \times \sqrt{3}$ phase is found at $J_2 = -0.07(1)$ for $J_2=0$. These phase boundaries determined in both ways are self-consistent and align well with previous studies~\cite{Gong2015, Kolley2015}.

\textit{Connection to kagome materials.--} 
Having established the ground state phase diagram of the spin-1/2 Heisenberg $J_1$-$J_2$-$D$ model on the kagome lattice, we then try to refine the parameters of the microscopic Hamiltonian for Zn-barlowite and herbertsmithite. This is achieved by comparing the static spin structure factor $S(\mathbf{q})$ from DMRG simulation with the neutron scattering experiments.
Experimentally, the value of $S(\mathbf{q})$ is approximated by integrating $S_{mag} (\mathbf{q},\omega)$ in the frequency region $\omega=2.5$~meV and 6~meV,
which arise from scattering from the intrinsic kagome moments~\cite{Han2016, Breidenbach2025}.
To quantify the goodness of fit, we compare the quantity:
\begin{eqnarray}
\delta R(J_2,D)=\sqrt{\sum_{i=1}^3(R_i(J_2,D)-R_i^E)^2},
\label{Eq:NeuSca}
\end{eqnarray}
which represents the deviation between the DMRG simulations on the model Hamiltonian in Eq.~(\ref{Eq:Ham}) and the neutron data. The ratios of the structure factor $S(\mathbf{q})$ integrated around several high-symmetry momenta are selected. These include $R_1=S({\rm \Gamma'})/S({\rm K'})$, $R_2=S({\rm \Gamma'})/S(\rm G)$ and $R_3=S({\rm K'})/S(\rm P)$, as defined in Fig.~\ref{Fig:latt_and_phd}(b). Here, $R_i$ and $R_i^E$ are the ratios from DMRG simulations and neutron experiments, respectively.

\begin{figure}
  \includegraphics[width=\linewidth]{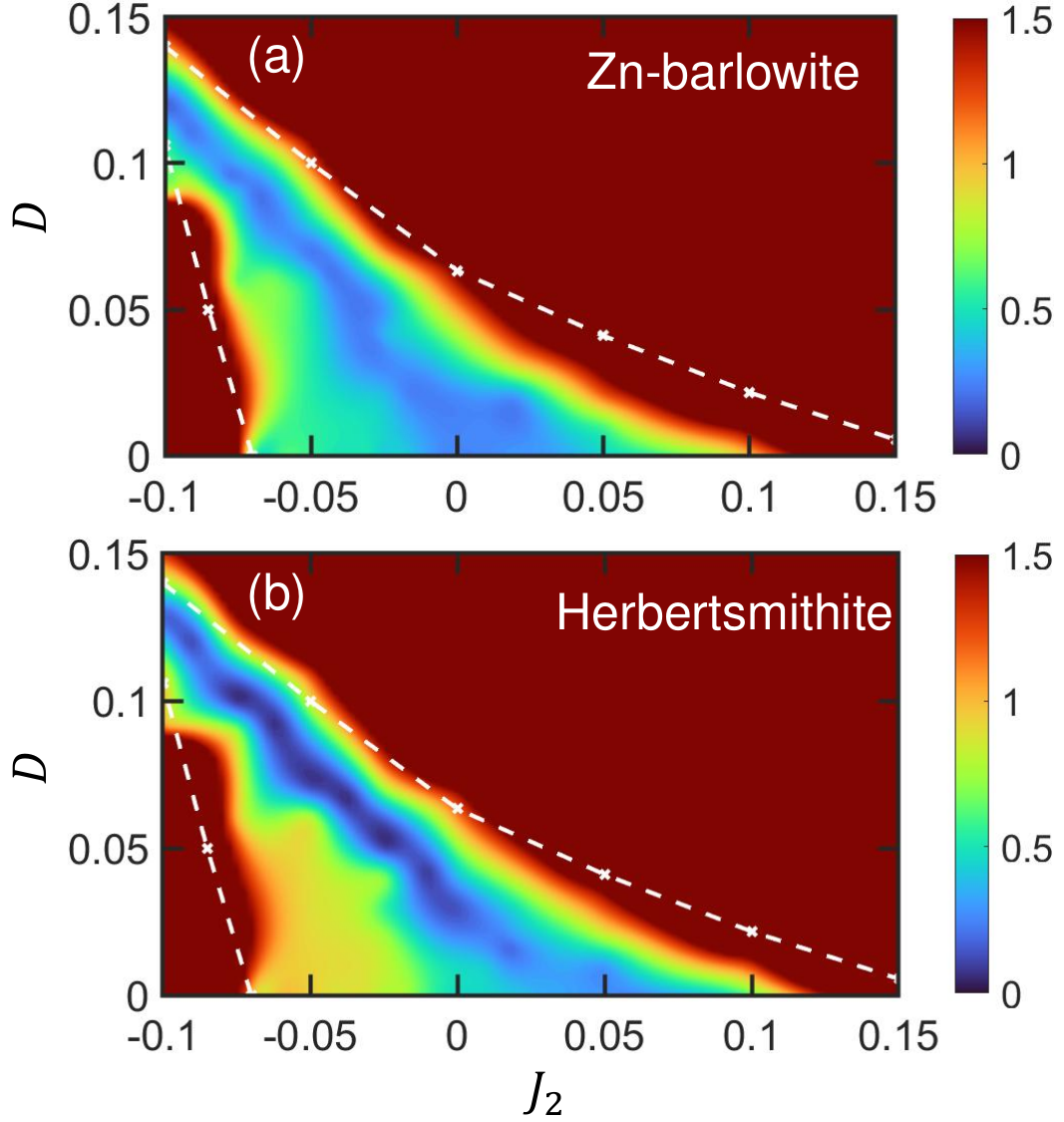}
  \caption{Contour plot of the deviation $\delta R(J_2,D)$ (defined in Eq.~\ref{Eq:NeuSca}) between DMRG simulations as a function of $J_2$ and $D$ on XC12 cylinders and neutron scattering data on (a) Zn-barlowite~\cite{Breidenbach2025} and (b) herbertsmithite~\cite{Han2012,HBS_unpublished}. The white dashed lines denote phase boundaries in Fig.~\ref{Fig:latt_and_phd}. The best-fit regions of both materials fall within the QSL phase. The contour is plotted with interpolation and an upper cutoff of 1.5 in the color scale.}
  \label{Fig:RContour}
\end{figure}

Figure \ref{Fig:RContour} shows the $\delta R(J_2,D)$ between DMRG simulation on XC12 kagome cylinders
and the neutron measurement on Zn-barlowite ($R_1$=1.126, $R_2$=1.430, $R_3$=1.130)~\cite{Breidenbach2025} as well as herbertsmithite ($R_1$=1.41, $R_2$=1.61, $R_3$=0.96)~\cite{Han2012,HBS_unpublished}. There is a boundary at $\delta R\approx 1.5$ beyond which $\delta R$ increases rapidly, enabling us to constrain the exchange parameters within this region, as shown in the supplementary information~\cite{sm}. Within the boundaries, $\delta R$ is significantly smaller and flatter. Moreover, experimental uncertainties, such as the energy integration window, can influence the details of this region. For instance, one would ideally integrate over all frequencies in the INS data. However, here we chose the frequency window of $\omega=[2.5, 6]$~meV, where non-intrinsic scattering from impurities and non-magnetic phonon scattering is minimized, as well as allowing for sufficient kinematic coverage in reciprocal space for integration~\cite{Han2016, Breidenbach2025}. 
From the simulation point of view, we always observe a finite spin triplet gap on the finite cylinders we study, although the gap can vary depending on the parameters and cylinder width. For example, on an XC8-12 cylinder, the triplet gap is $\sim 0.15J_1$ for ($J_2$=0, $D$=0), and $\sim 0.12J_1$ for ($J_2$=-0.02$J_1$, $D$=0), consistent with previous DMRG studies~\cite{Depenbrock2012,Kolley2015}. Using a $J_1$=15meV, there is no spectral weight within this $\sim2$meV gap, justifying using a lower cutoff in the energy integration for the INS.
Additionally, we note that the boundaries depicted in the figure are quite sharp and are relatively insensitive to the exact range of integration. For example, when enlarging the frequency integration window to $\omega=[2, 6.5]$~meV, the average change in each ratio is only $\sim 0.03$ for Zn-barlowite and $\sim0.01$ for herbertsmithite. Since the boundaries occur where $\delta R$ undergoes an abrupt change of order one, they are relatively unaffected. See supplementary information~\cite{sm} for more details.
Taking all these factors into account, the comparison of $S(\mathbf{q})$ between DMRG and neutron scattering allows us to constrain the exchange parameters to the region $\delta R\lesssim 1$ as shown in Fig.~\ref{Fig:RContour}, although it does not pinpoint one specific parameter set.
While our constraints do not provide an upper bound on $D$, electron spin resonance measurements suggest $D\approx0.08J_1$~\cite{Zorko2008} in herbertsmithite, hence we plot the range of $D$ to approximately twice this value in Fig.~\ref{Fig:RContour}.
Within the uncertainties, the constrained region from this quantitative structure factor comparison lies entirely within the QSL phase determined by DMRG simulations.
These comparisons provide strong additional evidence supporting the presence of the QSL state in both kagome materials.

\begin{figure}[t]
  \includegraphics[width=\linewidth]{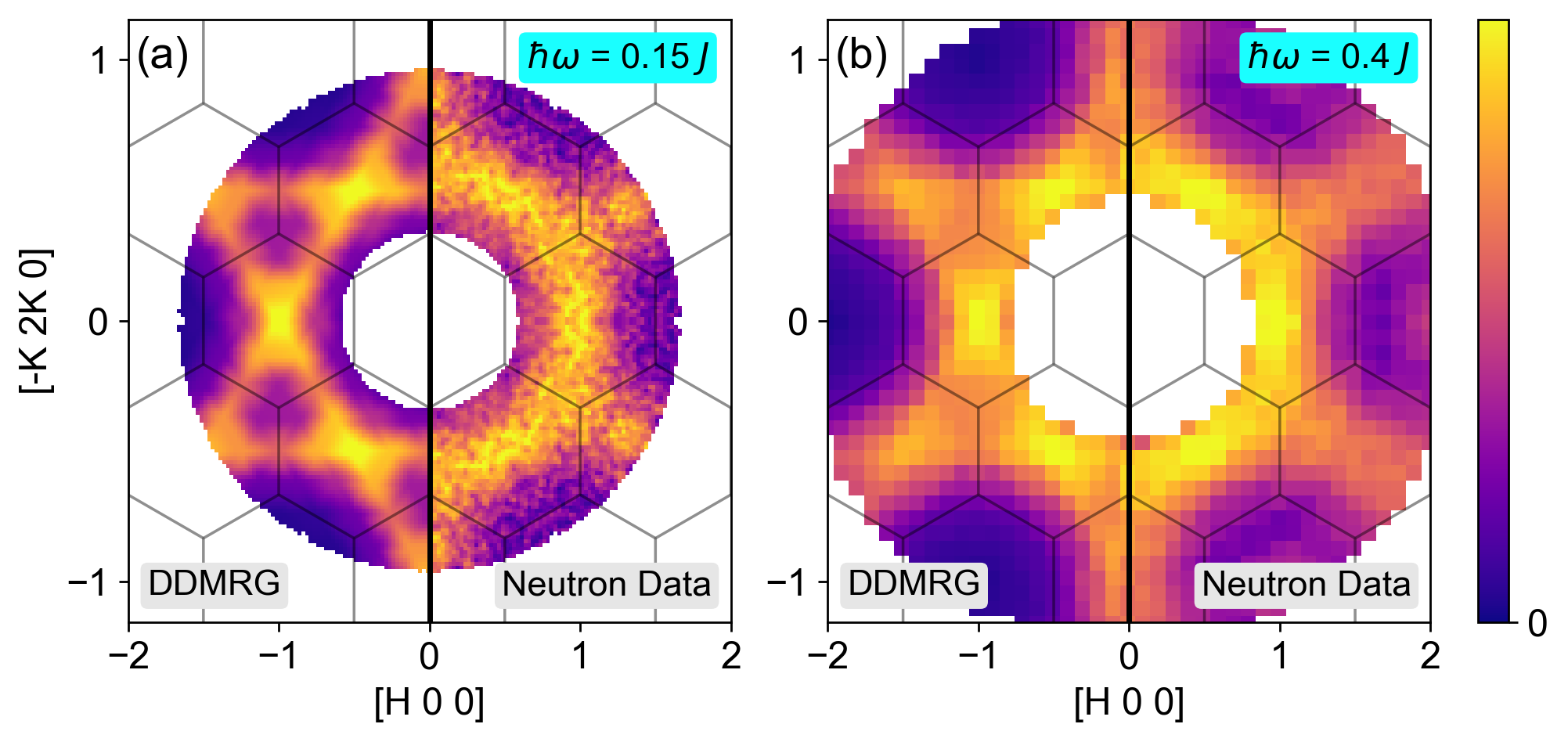}
  \caption{The dynamic spin structure factor $S(\mathbf{q},\omega)$, comparing results from DDMRG simulations (form factor adjusted) and measured inelastic neutron scattering data in the $(HK0)$ plane for Zn-barlowite~\cite{Breidenbach2025} for (a) $\hbar\omega=0.15J$ and (b) $\hbar\omega=0.4J$  at $T=1.7$~K. For the comparisons, the results are $D_6$-symmetrized and shown on relative intensity scales with the structural Brillouin zones overlaid.}\label{Fig:ZnbCompare}
\end{figure}

To further bridge the model with experiments, we have performed dynamical-DMRG (DDMRG)~\cite{ddmrg-jeckelmann,Zhu2019} simulation and obtained the dynamic spin structure factor $S(\mathbf{q},\omega)$, which we directly compared with recent inelastic neutron scattering measurements on Zn barlowite~\cite{Breidenbach2025}. 
The DDMRG simulations are carried out on XC8-11 cylinder with one additional edge column to reduce the boundary effect, using a broadening factor $\eta=0.1J_1$ and keeping a maximum bond dimension of 2400. More details of DDMRG are provided in supplementary information~\cite{sm}.
We choose two characteristic energy slices $\hbar\omega=0.15J_1$ and $\hbar\omega=0.4J_1$, and compared the resulting $S(\mathbf{q},\omega)$ between DDMRG simulation using ($J_2=-0.02J_1$, $D=0$) and INS (using $J_1=15$~meV) in Fig.~\ref{Fig:ZnbCompare}. 
The agreement between DDMRG simulations (symmetrized and magnetic form factor adjusted) and neutron scattering data is excellent, especially for the higher energy slice ($\hbar\omega=0.4J_1$). 
The slight discrepancy near $\rm K^\prime$ for $\hbar\omega=0.15J_1$ may stem from finite-size effects of the XC8 cylinder that frustrates the $\sqrt{3}\times\sqrt3$ order, which contributes to the intensity at the $\rm K'$ point, as well as non-intrinsic scattering from interlayer impurities which exist at low energies~\cite{Breidenbach2025}. We have also examined the pure Heisenberg model ($J_1$ only, see supplementary information~\cite{sm}), which shows a stronger peak at $\rm \Gamma'$ point with slightly poorer agreement with the neutron data. This suggests that a small ferromagnetic $J_2$ is likely present in Zn-barlowite, consistent with a previous analysis~\cite{Breidenbach2025}. A more precise determination of the parameters will require multiple simulations with different parameter sets $(J_2, D)$, which we will explore in future work.

\textit{Summary: }
We performed large-scale DMRG simulations of the kagome antiferromagnet, incorporating both the NNN exchange $J_2$ and DM interaction $D$, a model of direct relevance to real kagome materials. Through extensive simulations on 2D-like lattice geometries, we have identified an extended QSL phase in the ground-state phase diagram and established its robustness across a broad parameter space of $J_2$ and $D$. By comparing both static and dynamic spin structure factors from DMRG simulations with inelastic neutron scattering data, we further refined the exchange parameters for Zn-barlowite and herbertsmithite. 
Our findings reveal that the microscopic models for both materials reside within the QSL phase established by DMRG simulations, providing additional support for the presence of QSL ground states in these kagome compounds. 
Continued experimental advances in inelastic neutron scattering on single crystals, together with the development of large-scale DDMRG simulations incorporating nonzero $J_2$, $D$, and exchange anisotropy~\cite{Han2012a} terms, will enable a more precise determination of the model Hamiltonians for these QSL candidates, which is key to validating leading theoretical ideas.

\textit{Data availability}: 
The data used to generate the figures are deposited in Ref.~\footnote{\href{https://doi.org/10.6084/m9.figshare.28755095}{doi.org/10.6084/m9.figshare.28755095}}.

\textit{Acknowledgment:}
We acknowledge helpful discussions with Steve White regarding DDMRG. We acknowledge Aaron Breidenbach for sample synthesis, preparation, and prior measurements of Zn-barlowite. This work was supported by the U.S. Department of Energy (DOE), Office of Science, Basic Energy Sciences, Materials Sciences and Engineering Division, under contract DE-AC02-76SF00515. S.J. and the DDMRG calculations were supported by the Department of Energy, Laboratory Directed Research and Development program at SLAC National Accelerator Laboratory, under contract DE-AC02-76SF00515, with partial support for H.C.J.. This research used resources at the Spallation Neutron Source, Department of Energy (DOE), Office of Science User Facilities, operated by the Oak Ridge National Laboratory. DDMRG calculations were performed on resources of the National Energy Research Scientific Computing Center, supported by the U.S. Department of Energy under contract DE-AC02-05CH11231.

\bibliography{Refs}

@article{Anderson1973,
title = "Resonating valence bonds: A new kind of insulator?",
journal = "Mater. Res. Bull.",
volume = "8",
number = "2",
pages = "153 - 160",
year = "1973",
issn = "0025-5408",
doi = "https://doi.org/10.1016/0025-5408(73)90167-0",
author = "P.W. Anderson"
}

@article{Balents2010,
  title={Spin liquids in frustrated magnets},
  author={Balents, L.},
  journal={Nature},
  volume={464},
  number={7286},
  pages={199--208},
  year={2010},
  publisher={Nature Publishing Group},
  doi={10.1038/nature08917}
}

@article {Broholm2019,
	author = {Broholm, C. and Cava, R. J. and Kivelson, S. A. and Nocera, D. G. and Norman, M. R. and Senthil, T.},
	title = {Quantum spin liquids},
	volume = {367},
	number = {6475},
	pages = {eaay0668},
	elocation-id = {eaay0668},
	year = {2020},
	doi = {10.1126/science.aay0668},
	publisher = {American Association for the Advancement of Science},
	journal = {Science}
}

@article{Cepas2008,
  title = {Quantum phase transition induced by Dzyaloshinskii-Moriya interactions in the kagome antiferromagnet},
  author = {C\'epas, O. and Fong, C. M. and Leung, P. W. and Lhuillier, C.},
  journal = {Phys. Rev. B},
  volume = {78},
  issue = {14},
  pages = {140405},
  numpages = {4},
  year = {2008},
  month = {Oct},
  publisher = {American Physical Society},
  doi = {10.1103/PhysRevB.78.140405},
  url = {https://link.aps.org/doi/10.1103/PhysRevB.78.140405}
}

@article{Depenbrock2012,
  title = {Nature of the Spin-Liquid Ground State of the $S=1/2$ Heisenberg Model on the Kagome Lattice},
  author = {Depenbrock, S. and McCulloch, I. P. and Schollw\"ock, U.},
  journal = {Phys. Rev. Lett.},
  volume = {109},
  issue = {6},
  pages = {067201},
  numpages = {6},
  year = {2012},
  month = {Aug},
  publisher = {American Physical Society},
  doi = {10.1103/PhysRevLett.109.067201}
}

@article{Dzyaloshinsky1958,
title = "A thermodynamic theory of ``weak'' ferromagnetism of antiferromagnetics",
journal = "J. Phys. Chem. Solids. ",
volume = "4",
number = "4",
pages = "241 - 255",
year = "1958",
issn = "0022-3697",
doi = "https://doi.org/10.1016/0022-3697(58)90076-3",
url = "http://www.sciencedirect.com/science/article/pii/0022369758900763",
author = "I. Dzyaloshinsky"
}

@article{Elhajal2002,
  title = {Symmetry breaking due to Dzyaloshinsky-Moriya interactions in the kagom\'e lattice},
  author = {Elhajal, M. and Canals, B. and Lacroix, C.},
  journal = {Phys. Rev. B},
  volume = {66},
  issue = {1},
  pages = {014422},
  numpages = {6},
  year = {2002},
  month = {Jul},
  publisher = {American Physical Society},
  doi = {10.1103/PhysRevB.66.014422},
  url = {https://link.aps.org/doi/10.1103/PhysRevB.66.014422}
}

@article{Feng2017,
	doi = {10.1088/0256-307x/34/7/077502},
	url = {https://doi.org/10.1088%2F0256-307x%2F34%2F7%2F077502},
	year = 2017,
	month = {jun},
	publisher = {{IOP} Publishing},
	volume = {34},
	number = {7},
	pages = {077502},
	author = {Zili Feng and Zheng Li and Xin Meng and Wei Yi and Yuan Wei and Jun Zhang and Yan-Cheng Wang and Wei Jiang and Zheng Liu and Shiyan Li and Feng Liu and Jianlin Luo and Shiliang Li and Guo-qing Zheng and Zi Yang Meng and Jia-Wei Mei and Youguo Shi},
	title = "{Gapped Spin-1/2 Spinon Excitations in a New Kagome Quantum Spin Liquid Compound Cu$_3$Zn(OH)$_6$FBr}",
	journal = {Chin. Phys. Lett.}
}

@article {Fu2015,
	author = {Fu, Mingxuan and Imai, Takashi and Han, Tian-Heng and Lee, Young S.},
	title = {Evidence for a gapped spin-liquid ground state in a kagome Heisenberg antiferromagnet},
	volume = {350},
	number = {6261},
	pages = {655--658},
	year = {2015},
	publisher = {American Association for the Advancement of Science},
	journal = {Science},
    doi={10.1126/science.aab2120}
}

@article{Zhu2019,
author = {W. Zhu  and Shou-shu Gong  and D. N. Sheng },
title = {Identifying spinon excitations from dynamic structure factor of spin-1/2 Heisenberg antiferromagnet on the Kagome lattice},
journal = {Proc. Natl. Acad. Sci. U. S. A.},
volume = {116},
number = {12},
pages = {5437-5441},
year = {2019},
doi = {10.1073/pnas.1807840116},
}

@article {Han2012,
	author = {Han, T. H. and Helton, J. S. and Chu, S. and Nocera, D. G. and Rodriguez-Rivera, J. A. and Broholm, C. and Lee, Y. S.},
	title = {Fractionalized Excitations in the Spin-Liquid State of a Kagome Lattice Antiferromagnet},
	volume = {132},
	pages = {5570},
	year = {2012},
	publisher={Nature Publishing Group},
	journal = {Nature},
    doi={10.1038/nature11659}
}

@article{Gong2015,
  title = {Global phase diagram of competing ordered and quantum spin-liquid phases on the kagome lattice},
  author = {Gong, Shou-Shu and Zhu, Wei and Balents, Leon and Sheng, D. N.},
  journal = {Phys. Rev. B},
  volume = {91},
  issue = {7},
  pages = {075112},
  numpages = {9},
  year = {2015},
  month = {Feb},
  publisher = {American Physical Society},
  doi = {10.1103/PhysRevB.91.075112},
  url = {https://link.aps.org/doi/10.1103/PhysRevB.91.075112}
}

@article{Han2012a,
  title = "{Refining the Spin Hamiltonian in the Spin-$\frac{1}{2}$ Kagome Lattice Antiferromagnet ${\mathrm{ZnCu}}_{3}(\mathrm{OH}{)}_{6}{\mathrm{Cl}}_{2}$ Using Single Crystals}",
  author = {Han, Tianheng and Chu, Shaoyan and Lee, Young S.},
  journal = {Phys. Rev. Lett.},
  volume = {108},
  issue = {15},
  pages = {157202},
  numpages = {5},
  year = {2012},
  month = {Apr},
  publisher = {American Physical Society},
  doi = {10.1103/PhysRevLett.108.157202},
  url = {https://link.aps.org/doi/10.1103/PhysRevLett.108.157202}
}

@article{Han2016,
  title = {Correlated impurities and intrinsic spin-liquid physics in the kagome material herbertsmithite},
  author = {Han, Tian-Heng and Norman, M. R. and Wen, J.-J. and Rodriguez-Rivera, Jose A. and Helton, Joel S. and Broholm, Collin and Lee, Young S.},
  journal = {Phys. Rev. B},
  volume = {94},
  issue = {6},
  pages = {060409},
  numpages = {5},
  year = {2016},
  month = {Aug},
  publisher = {American Physical Society},
  doi = {10.1103/PhysRevB.94.060409},
  url = {https://link.aps.org/doi/10.1103/PhysRevB.94.060409}
}

@article{He2017,
  title = {Signatures of Dirac Cones in a DMRG Study of the Kagome Heisenberg Model},
  author = {He, Yin-Chen and Zaletel, Michael P. and Oshikawa, Masaki and Pollmann, Frank},
  journal = {Phys. Rev. X},
  volume = {7},
  issue = {3},
  pages = {031020},
  numpages = {16},
  year = {2017},
  month = {Jul},
  publisher = {American Physical Society},
  doi = {10.1103/PhysRevX.7.031020},
  url = {https://link.aps.org/doi/10.1103/PhysRevX.7.031020}
}

@article{Helton2007,
  title = "{Spin Dynamics of the Spin-$1/2$ Kagome Lattice Antiferromagnet ${\mathrm{ZnCu}}_{3}(\mathrm{OH}{)}_{6}{\mathrm{Cl}}_{2}$}",
  author = {Helton, J. S. and Matan, K. and Shores, M. P. and Nytko, E. A. and Bartlett, B. M. and Yoshida, Y. and Takano, Y. and Suslov, A. and Qiu, Y. and Chung, J.-H. and Nocera, D. G. and Lee, Y. S.},
  journal = {Phys. Rev. Lett.},
  volume = {98},
  issue = {10},
  pages = {107204},
  numpages = {4},
  year = {2007},
  month = {Mar},
  publisher = {American Physical Society},
  doi = {10.1103/PhysRevLett.98.107204},
  url = {https://link.aps.org/doi/10.1103/PhysRevLett.98.107204}
}

@article{Jeschke2013,
  title = "{First-principles determination of Heisenberg Hamiltonian parameters for the spin-$\frac{1}{2}$ kagome antiferromagnet ZnCu${}_{3}$(OH)${}_{6}$Cl${}_{2}$}",
  author = {Jeschke, Harald O. and Salvat-Pujol, Francesc and Valent\'{\i}, Roser},
  journal = {Phys. Rev. B},
  volume = {88},
  issue = {7},
  pages = {075106},
  numpages = {5},
  year = {2013},
  month = {Aug},
  publisher = {American Physical Society},
  doi = {10.1103/PhysRevB.88.075106}
}

@article{Jiang2008,
  title = {Density Matrix Renormalization Group Numerical Study of the Kagome Antiferromagnet},
  author = {Jiang, H. C. and Weng, Z. Y. and Sheng, D. N.},
  journal = {Phys. Rev. Lett.},
  volume = {101},
  issue = {11},
  pages = {117203},
  numpages = {4},
  year = {2008},
  month = {Sep},
  publisher = {American Physical Society},
  doi = {10.1103/PhysRevLett.101.117203},
}

@article{Jiang2012,
  title={ Identifying Topological Order by Entanglement Entropy},
  author={Jiang, H. C. and Wang, Z. and Balents, L.},
  journal={Nat. Phys.},
  volume={8},
  pages={902--905},
  year={2012},
  publisher={Nature Publishing Group},
  doi={10.1038/nphys2465},
}

@article{Kivelson1987,
  title = {Topology of the resonating valence-bond state: Solitons and high-${T}_{c}$ superconductivity},
  author = {Kivelson, Steven A. and Rokhsar, Daniel S. and Sethna, James P.},
  journal = {Phys. Rev. B},
  volume = {35},
  issue = {16},
  pages = {8865--8868},
  numpages = {0},
  year = {1987},
  month = {Jun},
  publisher = {American Physical Society},
  doi={10.1103/PhysRevB.35.8865}
}

@article{Kivelson1989,
  title = {Statistics of holons in the quantum hard-core dimer gas},
  author = {Kivelson, Steven},
  journal = {Phys. Rev. B},
  volume = {39},
  issue = {1},
  pages = {259--264},
  numpages = {0},
  year = {1989},
  month = {Jan},
  publisher = {American Physical Society},
  doi = {10.1103/PhysRevB.39.259},
  url = {https://link.aps.org/doi/10.1103/PhysRevB.39.259}
}

@article{Lauchli2011,
  title = {Ground-state energy and spin gap of spin-$\frac{1}{2}$ Kagom\'e-Heisenberg antiferromagnetic clusters: Large-scale exact diagonalization results},
  author = {L\"auchli, Andreas M. and Sudan, Julien and S\o{}rensen, Erik S.},
  journal = {Phys. Rev. B},
  volume = {83},
  issue = {21},
  pages = {212401},
  numpages = {4},
  year = {2011},
  month = {Jun},
  publisher = {American Physical Society},
  doi = {10.1103/PhysRevB.83.212401},
  url = {https://link.aps.org/doi/10.1103/PhysRevB.83.212401}
}

@article{Lauchli2019,
  title = {$S=\frac{1}{2}$ kagome Heisenberg antiferromagnet revisited},
  author = {L\"auchli, Andreas M. and Sudan, Julien and Moessner, Roderich},
  journal = {Phys. Rev. B},
  volume = {100},
  issue = {15},
  pages = {155142},
  numpages = {7},
  year = {2019},
  month = {Oct},
  publisher = {American Physical Society},
  doi = {10.1103/PhysRevB.100.155142},
  url = {https://link.aps.org/doi/10.1103/PhysRevB.100.155142}
}

@article{Kolley2015,
  title = {Phase diagram of the ${J}_{1}\text{\ensuremath{-}}{J}_{2}$ Heisenberg model on the kagome lattice},
  author = {Kolley, F. and Depenbrock, S. and McCulloch, I. P. and Schollw\"ock, U. and Alba, V.},
  journal = {Phys. Rev. B},
  volume = {91},
  issue = {10},
  pages = {104418},
  numpages = {8},
  year = {2015},
  month = {Mar},
  publisher = {American Physical Society},
  doi = {10.1103/PhysRevB.91.104418},
}

@article{Laughlin1988,
  title = {Superconducting Ground State of Noninteracting Particles Obeying Fractional Statistics},
  author = {Laughlin, R. B.},
  journal = {Phys. Rev. Lett.},
  volume = {60},
  issue = {25},
  pages = {2677--2680},
  numpages = {0},
  year = {1988},
  month = {Jun},
  publisher = {American Physical Society},
  doi = {10.1103/PhysRevLett.60.2677},
  url = {http://link.aps.org/doi/10.1103/PhysRevLett.60.2677}
}

@article{Lee2013,
  title = {Proposal to use neutron scattering to access scalar spin chirality fluctuations in kagome lattices},
  author = {Lee, Patrick A. and Nagaosa, Naoto},
  journal = {Phys. Rev. B},
  volume = {87},
  issue = {6},
  pages = {064423},
  numpages = {4},
  year = {2013},
  month = {Feb},
  publisher = {American Physical Society},
  doi = {10.1103/PhysRevB.87.064423},
  url = {https://link.aps.org/doi/10.1103/PhysRevB.87.064423}
}

@article{Liao2017,
  title = {Gapless Spin-Liquid Ground State in the $S=1/2$ Kagome Antiferromagnet},
  author = {Liao, H. J. and Xie, Z. Y. and Chen, J. and Liu, Z. Y. and Xie, H. D. and Huang, R. Z. and Normand, B. and Xiang, T.},
  journal = {Phys. Rev. Lett.},
  volume = {118},
  issue = {13},
  pages = {137202},
  numpages = {6},
  year = {2017},
  month = {Mar},
  publisher = {American Physical Society},
  doi = {10.1103/PhysRevLett.118.137202},
  url = {https://link.aps.org/doi/10.1103/PhysRevLett.118.137202}
}

@article{Mei2017,
  title = {Gapped spin liquid with ${\mathbb{Z}}_{2}$ topological order for the kagome Heisenberg model},
  author = {Mei, Jia-Wei and Chen, Ji-Yao and He, Huan and Wen, Xiao-Gang},
  journal = {Phys. Rev. B},
  volume = {95},
  issue = {23},
  pages = {235107},
  numpages = {9},
  year = {2017},
  month = {Jun},
  publisher = {American Physical Society},
  doi = {10.1103/PhysRevB.95.235107},
  url = {https://link.aps.org/doi/10.1103/PhysRevB.95.235107}
}

@article{Mendels2007,
  title = {Quantum Magnetism in the Paratacamite Family: Towards an Ideal Kagom\'e Lattice},
  author = {Mendels, P. and Bert, F. and de Vries, M. A. and Olariu, A. and Harrison, A. and Duc, F. and Trombe, J. C. and Lord, J. S. and Amato, A. and Baines, C.},
  journal = {Phys. Rev. Lett.},
  volume = {98},
  issue = {7},
  pages = {077204},
  numpages = {4},
  year = {2007},
  month = {Feb},
  publisher = {American Physical Society},
  doi = {10.1103/PhysRevLett.98.077204},
  url = {https://link.aps.org/doi/10.1103/PhysRevLett.98.077204}
}

@article{Messio2010,
  title = {Schwinger-boson approach to the kagome antiferromagnet with Dzyaloshinskii-Moriya interactions: Phase diagram and dynamical structure factors},
  author = {Messio, L. and C\'epas, O. and Lhuillier, C.},
  journal = {Phys. Rev. B},
  volume = {81},
  issue = {6},
  pages = {064428},
  numpages = {8},
  year = {2010},
  month = {Feb},
  publisher = {American Physical Society},
  doi = {10.1103/PhysRevB.81.064428},
  url = {https://link.aps.org/doi/10.1103/PhysRevB.81.064428}
}

@article{Moriya1960,
  title = {Anisotropic Superexchange Interaction and Weak Ferromagnetism},
  author = {Moriya, T\^oru},
  journal = {Phys. Rev.},
  volume = {120},
  issue = {1},
  pages = {91--98},
  numpages = {0},
  year = {1960},
  month = {Oct},
  publisher = {American Physical Society},
  doi = {10.1103/PhysRev.120.91},
  url = {https://link.aps.org/doi/10.1103/PhysRev.120.91}
}

@article{Moessner2001,
  title = {Resonating Valence Bond Phase in the Triangular Lattice Quantum Dimer Model},
  author = {Moessner, R. and Sondhi, S. L.},
  journal = {Phys. Rev. Lett.},
  volume = {86},
  issue = {9},
  pages = {1881--1884},
  numpages = {0},
  year = {2001},
  month = {Feb},
  publisher = {American Physical Society},
  doi={10.1103/PhysRevLett.86.1881}
}

@article{Ran2007,
  title = {Projected-Wave-Function Study of the Spin-$1/2$ Heisenberg Model on the Kagom\'e Lattice},
  author = {Ran, Ying and Hermele, Michael and Lee, Patrick A. and Wen, Xiao-Gang},
  journal = {Phys. Rev. Lett.},
  volume = {98},
  issue = {11},
  pages = {117205},
  numpages = {4},
  year = {2007},
  month = {Mar},
  publisher = {American Physical Society},
  doi = {10.1103/PhysRevLett.98.117205},
  url = {https://link.aps.org/doi/10.1103/PhysRevLett.98.117205}
}

@article{Rigol2007,
  title = "{Magnetic Susceptibility of the Kagome Antiferromagnet ${\mathrm{ZnCu}}_{3}(\mathrm{OH}{)}_{6}{\mathrm{Cl}}_{2}$}",
  author = {Rigol, Marcos and Singh, Rajiv R. P.},
  journal = {Phys. Rev. Lett.},
  volume = {98},
  issue = {20},
  pages = {207204},
  numpages = {4},
  year = {2007},
  month = {May},
  publisher = {American Physical Society},
  doi = {10.1103/PhysRevLett.98.207204},
  url = {https://link.aps.org/doi/10.1103/PhysRevLett.98.207204}
}

@article{Rokhsar1988,
  title = {Superconductivity and the Quantum Hard-Core Dimer Gas},
  author = {Rokhsar, Daniel S. and Kivelson, Steven A.},
  journal = {Phys. Rev. Lett.},
  volume = {61},
  issue = {20},
  pages = {2376--2379},
  numpages = {0},
  year = {1988},
  month = {Nov},
  publisher = {American Physical Society},
  doi={10.1103/PhysRevLett.61.2376}
}

@article{Savary2016,
	doi = {10.1088/0034-4885/80/1/016502},
	year = 2016,
	month = {nov},
	publisher = {{IOP} Publishing},
	volume = {80},
	number = {1},
	pages = {016502},
	author = {Lucile Savary and Leon Balents},
	title = {Quantum spin liquids: a review},
	journal = {Rep. Prog. Phys.}
}

@article {Smaha2020,
  title={Materializing rival ground states in the barlowite family of kagome magnets: quantum spin liquid, spin ordered, and valence bond crystal states},
  author={Smaha, Rebecca W and He, Wei and Jiang, Jack Mingde and Wen, Jiajia and Jiang, Yi-Fan and Sheckelton, John P and Titus, Charles J and Wang, Suyin Grass and Chen, Yu-Sheng and Teat, Simon J and others},
  journal={npj Quantum Mater.},
  volume={5},
  number={1},
  pages={23},
  year={2020},
  publisher={Nature Publishing Group UK London},
  doi={10.1038/s41535-020-0230-8}
}

@article{Vries2009,
  title = {Scale-Free Antiferromagnetic Fluctuations in the $s=1/2$ Kagome Antiferromagnet Herbertsmithite},
  author = {de Vries, M. A. and Stewart, J. R. and Deen, P. P. and Piatek, J. O. and Nilsen, G. J. and R\o{}nnow, H. M. and Harrison, A.},
  journal = {Phys. Rev. Lett.},
  volume = {103},
  issue = {23},
  pages = {237201},
  numpages = {4},
  year = {2009},
  month = {Dec},
  publisher = {American Physical Society},
  doi = {10.1103/PhysRevLett.103.237201},
  url = {https://link.aps.org/doi/10.1103/PhysRevLett.103.237201}
}

@article{White1992,
  title = {Density matrix formulation for quantum renormalization groups},
  author = {White, Steven R.},
  journal = {Phys. Rev. Lett.},
  volume = {69},
  issue = {19},
  pages = {2863--2866},
  numpages = {0},
  year = {1992},
  month = {Nov},
  publisher = {American Physical Society},
  doi={10.1103/PhysRevLett.69.2863}
}

@article{White1993,
  title = {Density-matrix algorithms for quantum renormalization groups},
  author = {White, Steven R.},
  journal = {Phys. Rev. B},
  volume = {48},
  issue = {14},
  pages = {10345--10356},
  numpages = {0},
  year = {1993},
  month = {Oct},
  publisher = {American Physical Society},
  doi = {10.1103/PhysRevB.48.10345}
}

@article{Yan2011,
  title={Spin-liquid ground state of the s= 1/2 kagome heisenberg antiferromagnet},
  author={Yan, S. and Huse, D.A. and White, S.R.},
  journal={Science},
  volume={332},
  number={6034},
  pages={1173},
  year={2011},
  publisher={American Association for the Advancement of Science},
  doi={10.1126/science.1201080}
}

@article{Zorko2008,
  title = "{Dzyaloshinsky-Moriya Anisotropy in the Spin-1/2 Kagome Compound ${\mathrm{ZnCu}}_{3}(\mathrm{OH}{)}_{6}{\mathrm{Cl}}_{2}$}",
  author = {Zorko, A. and Nellutla, S. and van Tol, J. and Brunel, L. C. and Bert, F. and Duc, F. and Trombe, J.-C. and de Vries, M. A. and Harrison, A. and Mendels, P.},
  journal = {Phys. Rev. Lett.},
  volume = {101},
  issue = {2},
  pages = {026405},
  numpages = {4},
  year = {2008},
  month = {Jul},
  publisher = {American Physical Society},
  doi = {10.1103/PhysRevLett.101.026405},
  url = {https://link.aps.org/doi/10.1103/PhysRevLett.101.026405}
}

@article{Wang2021,
    author = {Wang, Jiaming and Yuan,Weishi and Singer,Philip M. and Smaha, Rebecca W. and He, Wei and Wen,Jiajia and Lee, Young S. and Imai, Takashi},
    title = {Emergence of spin singlets with inhomogeneous gaps in the kagome lattice Heisenberg antiferromagnets Zn-barlowite and herbertsmithite},
    journal = {Nat. Phys.},
    volume={17},
    pages={1109},
    year = {2021},
    doi={10.1038/s41567-021-01310-3}
}

@article{steve-sasha,
  title = {Ne\'el Order in Square and Triangular Lattice Heisenberg Models},
  author = {White, Steven R. and Chernyshev, A. L.},
  journal = {Phys. Rev. Lett.},
  volume = {99},
  issue = {12},
  pages = {127004},
  numpages = {4},
  year = {2007},
  month = {Sep},
  publisher = {American Physical Society},
  doi = {10.1103/PhysRevLett.99.127004},
  url = {https://link.aps.org/doi/10.1103/PhysRevLett.99.127004}
}

@article{Smaha2018,
title = {Synthesis-dependent properties of barlowite and Zn-substituted barlowite},
journal = {J. Solid State Chem},
volume = {268},
pages = {123-129},
year = {2018},
issn = {0022-4596},
doi = {https://doi.org/10.1016/j.jssc.2018.08.016},
url = {https://www.sciencedirect.com/science/article/pii/S0022459618303475},
author = {Rebecca W. Smaha and Wei He and John P. Sheckelton and Jiajia Wen and Young S. Lee},
}

@article{Tustain2020magnetic,
  title={From magnetic order to quantum disorder in the Zn-barlowite series of S= 1/2 kagom{\'e} antiferromagnets},
  author={Tustain, Katherine and Ward-O'Brien, Brendan and Bert, Fabrice and Han, Tianheng and Luetkens, Hubertus and Lancaster, Tom and Huddart, Benjamin M and Baker, Peter J and Clark, Lucy},
  journal={npj Quantum Materials},
  volume={5},
  number={1},
  pages={74},
  year={2020},
  publisher={Nature Publishing Group UK London},
  doi={10.1038/s41535-020-00276-4}
}

@article{Fu2021Dynamic,
  title={Dynamic fingerprint of fractionalized excitations in single-crystalline Cu$_3$Zn(OH)$_6$FBr},
  author={Fu, Ying and Lin, Miao-Ling and Wang, Le and Liu, Qiye and Huang, Lianglong and Jiang, Wenrui and Hao, Zhanyang and Liu, Cai and Zhang, Hu and Shi, Xingqiang and others},
  journal={Nat. Commun.},
  volume={12},
  number={1},
  pages={3048},
  year={2021},
  publisher={Nature Publishing Group UK London},
  doi = {10.1038/s41467-021-23381-9},
}

@article{Breidenbach2025,
  title = {Identifying Universal Spin Excitations in Spin-1/2 Kagome Quantum Spin Liquid Materials},
  author = {Breidenbach, Aaron T. and Campello, Arthur C. and Wen, Jiajia and Jiang, Hong-Chen and Pajerowski, Daniel M. and Smaha, Rebecca W. and Lee, Young S.},
  year = {2025},
  journal={arXiv:2504.06491},
  url={https://arxiv.org/abs/2504.06491}
}

@unpublished{HBS_unpublished,
  author = {Campello, Arthur C.},
  year = {2025},
  title = {},
  note={unpublished}
}

@article{ddmrg-jeckelmann,
  title = {Dynamical density-matrix renormalization-group method},
  author = {Jeckelmann, Eric},
  journal = {Phys. Rev. B},
  volume = {66},
  issue = {4},
  pages = {045114},
  numpages = {16},
  year = {2002},
  month = {Jul},
  publisher = {American Physical Society},
  doi = {10.1103/PhysRevB.66.045114},
  url = {https://link.aps.org/doi/10.1103/PhysRevB.66.045114}
}

@article{sun2024possible,
  title={Possible chiral spin liquid state in the S= 1/2 kagome Heisenberg model},
  author={Sun, Rong-Yang and Jin, Hui-Ke and Tu, Hong-Hao and Zhou, Yi},
  journal={npj Quantum Mater.},
  volume={9},
  number={1},
  pages={16},
  year={2024},
  publisher={Nature Publishing Group UK London},
  doi={10.1038/s41535-024-00627-5}
}

@article{Haghshenas2019,
  title = {Single-layer tensor network study of the Heisenberg model with chiral interactions on a kagome lattice},
  author = {Haghshenas, R. and Gong, Shou-Shu and Sheng, D. N.},
  journal = {Phys. Rev. B},
  volume = {99},
  issue = {17},
  pages = {174423},
  numpages = {11},
  year = {2019},
  month = {May},
  publisher = {American Physical Society},
  doi = {10.1103/PhysRevB.99.174423},
  url = {https://link.aps.org/doi/10.1103/PhysRevB.99.174423}
}

@article{ed1,
    doi = {10.1209/0295-5075/88/27009},
    url = {https://dx.doi.org/10.1209/0295-5075/88/27009},
    year = {2009},
    month = {nov},
    publisher = {},
    volume = {88},
    number = {2},
    pages = {27009},
    author = {Sindzingre, P. and Lhuillier, C.},
    title = {Low-energy excitations of the kagome antiferromagnet and the spin-gap issue},
    journal = {EPL}
}

@article{gapless-exp1,
  title="{Gapless ground state in the archetypal quantum kagome antiferromagnet ZnCu$_3$(OH)$_6$Cl$_2$}",
  author={Khuntia, P and Vel{\'a}zquez, Matias and Barth{\'e}lemy, Quentin and Bert, Fabrice and Kermarrec, Edwin and Legros, A and Bernu, Bernard and Messio, L and Zorko, Andrej and Mendels, P},
  journal={Nat. Phys.},
  volume={16},
  number={4},
  pages={469--474},
  year={2020},
  publisher={Nature Publishing Group UK London},
  doi={10.1038/s41567-020-0792-1}
}

@article{Norman2016,
  title = {Colloquium: Herbertsmithite and the search for the quantum spin liquid},
  author = {Norman, M. R.},
  journal = {Rev. Mod. Phys.},
  volume = {88},
  issue = {4},
  pages = {041002},
  numpages = {14},
  year = {2016},
  month = {Dec},
  publisher = {American Physical Society},
  doi = {10.1103/RevModPhys.88.041002},
  url = {https://link.aps.org/doi/10.1103/RevModPhys.88.041002}
}

@article{sasha2015,
  title = {Order and excitations in $\text{large}\ensuremath{-}S$ kagome-lattice antiferromagnets},
  author = {Chernyshev, A. L. and Zhitomirsky, M. E.},
  journal = {Phys. Rev. B},
  volume = {92},
  issue = {14},
  pages = {144415},
  numpages = {16},
  year = {2015},
  month = {Oct},
  publisher = {American Physical Society},
  doi = {10.1103/PhysRevB.92.144415},
}

@article{harris1992,
  title = {Possible N\'eel orderings of the Kagom\'e antiferromagnet},
  author = {Harris, A. B. and Kallin, C. and Berlinsky, A. J.},
  journal = {Phys. Rev. B},
  volume = {45},
  issue = {6},
  pages = {2899--2919},
  numpages = {0},
  year = {1992},
  month = {Feb},
  publisher = {American Physical Society},
  doi = {10.1103/PhysRevB.45.2899},
  url = {https://link.aps.org/doi/10.1103/PhysRevB.45.2899}
}

@article{chubukov1992,
  title = {Order from disorder in a kagom\'e antiferromagnet},
  author = {Chubukov, Andrey},
  journal = {Phys. Rev. Lett.},
  volume = {69},
  issue = {5},
  pages = {832--835},
  numpages = {0},
  year = {1992},
  month = {Aug},
  publisher = {American Physical Society},
  doi = {10.1103/PhysRevLett.69.832},
  url = {https://link.aps.org/doi/10.1103/PhysRevLett.69.832}
}

@article{huh2010,
  title = {Quantum criticality of the kagome antiferromagnet with Dzyaloshinskii-Moriya interactions},
  author = {Huh, Yejin and Fritz, Lars and Sachdev, Subir},
  journal = {Phys. Rev. B},
  volume = {81},
  issue = {14},
  pages = {144432},
  numpages = {8},
  year = {2010},
  month = {Apr},
  publisher = {American Physical Society},
  doi = {10.1103/PhysRevB.81.144432},
  url = {https://link.aps.org/doi/10.1103/PhysRevB.81.144432}
}

@article{vmc1,
  title = {Gutzwiller projected states for the ${J}_{1}\ensuremath{-}{J}_{2}$ Heisenberg model on the Kagome lattice: Achievements and pitfalls},
  author = {Iqbal, Yasir and Ferrari, Francesco and Chauhan, Aishwarya and Parola, Alberto and Poilblanc, Didier and Becca, Federico},
  journal = {Phys. Rev. B},
  volume = {104},
  issue = {14},
  pages = {144406},
  numpages = {9},
  year = {2021},
  month = {Oct},
  publisher = {American Physical Society},
  doi = {10.1103/PhysRevB.104.144406},
  url = {https://link.aps.org/doi/10.1103/PhysRevB.104.144406}
}

@Article{vmc2,
	title={{Static and dynamical signatures of Dzyaloshinskii-Moriya interactions in the Heisenberg model on the kagome lattice}},
	author={Francesco Ferrari and Sen Niu and Juraj Hasik and Yasir Iqbal and Didier Poilblanc and Federico Becca},
	journal={SciPost Phys.},
	volume={14},
	pages={139},
	year={2023},
	publisher={SciPost},
	doi={10.21468/SciPostPhys.14.6.139},
	url={https://scipost.org/10.21468/SciPostPhys.14.6.139},
}

@article{vmctaoli,
  title = "{Variational study of the ground state and spin dynamics of the spin-$\frac{1}{2}$ kagome antiferromagnetic Heisenberg model and its implication for herbertsmithite ${\mathrm{ZnCu}}_{3}{(\mathrm{OH})}_{6}{\mathrm{Cl}}_{2}$}",
  author = {Zhang, Chun and Li, Tao},
  journal = {Phys. Rev. B},
  volume = {102},
  issue = {19},
  pages = {195106},
  numpages = {11},
  year = {2020},
  month = {Nov},
  publisher = {American Physical Society},
  doi = {10.1103/PhysRevB.102.195106},
  url = {https://link.aps.org/doi/10.1103/PhysRevB.102.195106}
}

@misc{sm,
	note={See Supplemental Material at [], for the details on the DMRG calculations, additional results, and details of the analytical MAGSWT formalism.}
}

\end{document}


\title{Supplementary Information: Quantifying the Phase Diagram and Hamiltonian of $S=1/2$ Kagome Antiferromagnets: Bridging Theory and Experiment}

\begin{CJK*}{UTF8}{}
\author{Shengtao Jiang (\CJKfamily{gbsn}蒋晟韬)}
\affiliation{Stanford Institute for Materials and Energy Sciences, SLAC National Accelerator Laboratory, Menlo Park, CA 94025, USA}

\author{Arthur C. Campello}
\affiliation{Stanford Institute for Materials and Energy Sciences, SLAC National Accelerator Laboratory, Menlo Park, CA 94025, USA}
\affiliation{Department of Applied Physics, Stanford University, Stanford, CA 94305, USA}

\author{Wei He}
\affiliation{Stanford Institute for Materials and Energy Sciences, SLAC National Accelerator Laboratory, Menlo Park, CA 94025, USA}

\author{Jiajia Wen}
\affiliation{Stanford Institute for Materials and Energy Sciences, SLAC National Accelerator Laboratory, Menlo Park, CA 94025, USA}

\author{Daniel M. Pajerowski}
\affiliation{Neutron Scattering Division, Oak Ridge National Laboratory, Oak Ridge TN 37830, USA}

\author{Young S. Lee}
\affiliation{Stanford Institute for Materials and Energy Sciences, SLAC National Accelerator Laboratory, Menlo Park, CA 94025, USA}
\affiliation{Department of Applied Physics, Stanford University, Stanford, CA 94305, USA}

\author{Hong-Chen Jiang}
\email{hcjiang@stanford.edu}
\affiliation{Stanford Institute for Materials and Energy Sciences, SLAC National Accelerator Laboratory, Menlo Park, CA 94025, USA}
\date{\today}
\maketitle
\end{CJK*}

\section{Line plot of $\delta R(J_2,D)$ beyond cutoff}
\begin{figure}[h]
  \includegraphics[width=0.7\linewidth]{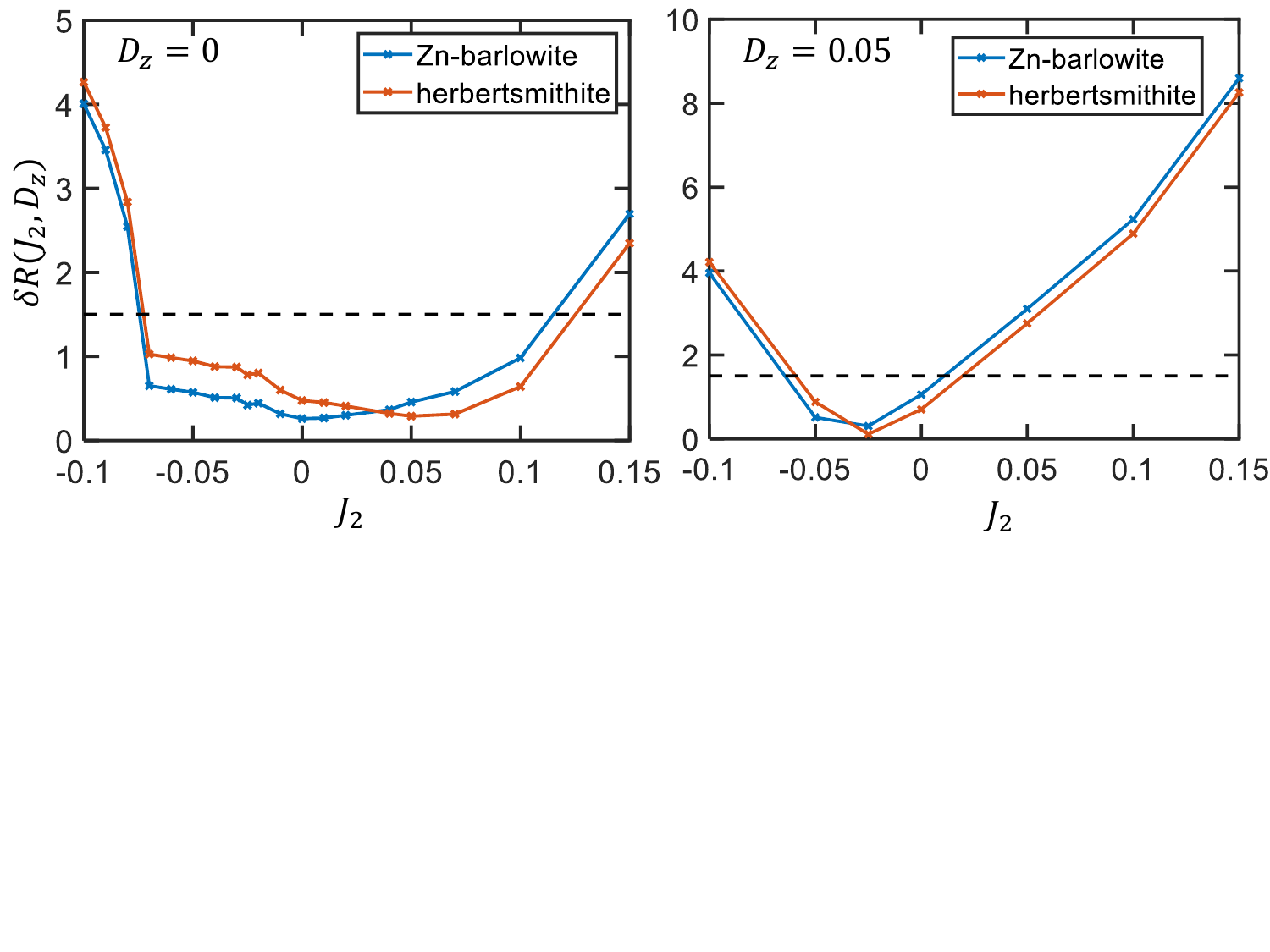}
  \vskip -4.0cm
  \caption{Line plot of $\delta R(J_2,D)$ at fixed $D=0$ and $D=0.05$ between DMRG simulation and INS data for both materials. The black dashed line is the cutoff of 1.5 used in Fig.~4 in the main text.}
  \label{Fig:deltaR_line}
\end{figure}

In the contour plots in Fig.~4 in the main text, we apply an upper cutoff of 1.5 on the color scale to emphasize the variation below this cutoff. Here in Fig.~\ref{Fig:deltaR_line} we show the line plot of $\delta R(J_2,D)$ along $D=0$ and $D=0.05$. As we can see, $\delta R$ is relatively flat below the cutoff but increases rapidly above the cutoff. This justifies constraining the $(J_2,D)$ for the materials to the region $\delta R(J_2,D)\lesssim1$, since the region outside has a clear deviation from experiments.

\section{Dynamical DMRG simulation}
Our goal is to calculate the dynamical spin structural factor, defined as:
\begin{equation}
\begin{split}
S(\mathbf{q,\omega})=&\frac{-1}{\pi N}\sum_{\substack{i,j=1\\\alpha=x,y,z}}^{N}e^{i\mathbf{q}\cdot(\mathbf{r}_i-\mathbf{r}_j)}\times\\
&\text{Im}\langle 0|S^\alpha_{\mathbf{r}_i}\frac{1}{E_0+\omega-H+i\eta}S^\alpha_{\mathbf{r}_j}|0\rangle,
\label{eq:dssf}
\end{split}
\end{equation}
where $|0\rangle$ is the ground state obtained by DMRG that has energy $E_0$, and $\eta$ is the broadening factor. 
We define the correction vector $|c^\alpha_{\mathbf{r}_i}(i,\omega)\rangle$ that satisfies:
\begin{equation}
(E_0+\omega-H+i\eta)|c^\alpha_{\mathbf{r}_j}(\omega)\rangle=S^\alpha_{\mathbf{r}_j}|0\rangle, 
\end{equation}
the $S(\mathbf{q,\omega})$ can then be computed with:
\begin{equation}
S(\mathbf{q,\omega})=\frac{-1}{\pi N}\sum_{\substack{i,j=1\\\alpha=x,y,z}}^{N}e^{i\mathbf{q}\cdot(\mathbf{r}_i-\mathbf{r}_j)}\times\text{Im}\langle 0|S^\alpha_{\mathbf{r}_i}|c^\alpha_{\mathbf{r}_j}(\omega)\rangle.
\end{equation}
The correction vector is separated into real and imaginary parts: $|X^\alpha_{\mathbf{r}_j}(\omega)\rangle=\text{Re}|c^\alpha_{\mathbf{r}_j}(\omega)\rangle$ and $|Y^\alpha_{\mathbf{r}_j}(\omega)\rangle=\text{Im}|c^\alpha_{\mathbf{r}_j}(\omega)\rangle$, so that it can be solved with real arithmetic. Its imaginary part satisfies:
\begin{equation}
    [(E_0+\omega-H)^2+\eta^2]|Y^\alpha_{\mathbf{r}_j}(\omega)\rangle=-\eta S^\alpha_{\mathbf{r}_j}|0\rangle, 
    \label{eq:ddmrg}
\end{equation}
based on which the real part can also be computed:
\begin{equation}
    |X^\alpha_{\mathbf{r}_j}(\omega)\rangle=\frac{H-E_0-\omega}{\eta}|Y^\alpha_{\mathbf{r}_j}(\omega)\rangle.
\end{equation}
The dynamical-DMRG algorithm~\cite{ddmrg-jeckelmann} solves Eq.~\ref{eq:ddmrg} using the same sweep scheme as the DMRG algorithm~\cite{White1992,White1993}, replacing the original eigenvalue problem with the linear problem in Eq.~\ref{eq:ddmrg}. At each step of the sweep, an effective superblock Hamiltonian $H^{\rm eff}$ is formed by projecting the original Hamiltonian on the reduced basis. $H^{\rm eff}$ approximates the $H$ in Eq.~\ref{eq:ddmrg}, and this linear equation is solved using the conjugate gradient method. To minimize the effect of this approximation, $|X^\alpha_{\mathbf{r}_j}(\omega)\rangle$ needs to be state-averaged with $|Y^\alpha_{\mathbf{r}_j}(\omega)\rangle$ in DDMRG. 

\begin{figure}[ht]
  \includegraphics[width=0.7\linewidth]{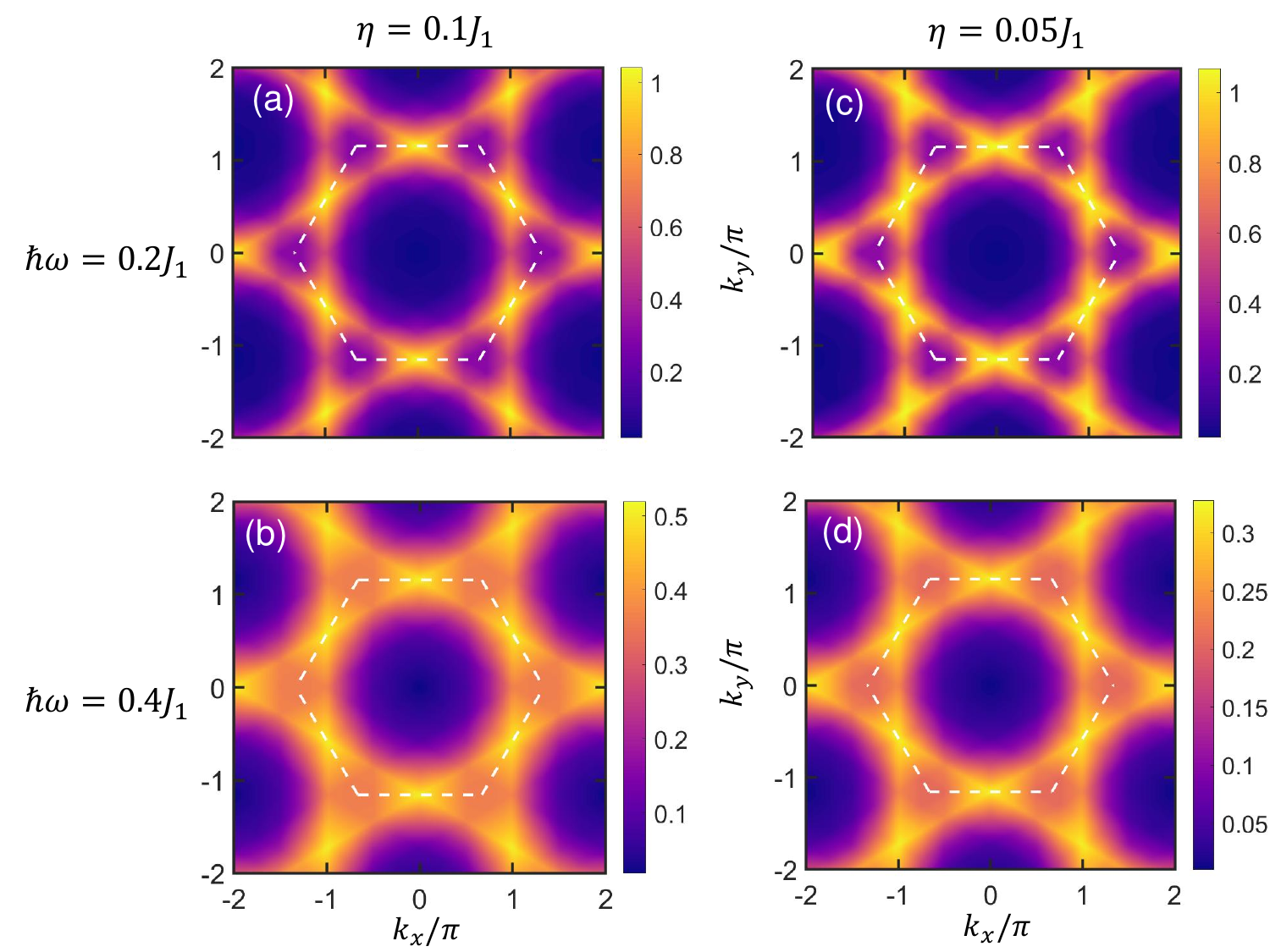}
  \caption{The $S(\mathbf{k},\omega)$ obtained by DDMRG with ($J_2=-0.02J_1$, $D=0$) at $\hbar\omega=0.2J_1$ (top) and $\hbar\omega=0.4J_1$ (bottom), (a)(b) with $\eta=0.1J_1$ and (c)(d) $\eta=0.05J_1$. The white dashed lines are the extended BZ boundaries. The results are $D_6$ symmetrized and interpolated.}
  \label{Fig:etacomparison}
\end{figure}

In this paper, we select an $\eta=0.1J_1$ as the broadening factor. We also compared our results with using a smaller $\eta=0.05J_1$ in Fig.~\ref{Fig:etacomparison} and found only very minor quantitative changes (except for the overall scale that can change with $\eta$), demonstrating that our results are robust with respect to the choice of $\eta$. Because the SU(2) symmetry is preserved in the QSL state we study, we replace the summation over spin component $\alpha$ in Eq.~\ref{eq:dssf} by the $z$ component multiplied by three.
To reduce boundary effects, the DDMRG simulations are carried out on XC8-11 cylinder with one additional edge column, and the Fourier transform in Eq.~\ref{eq:dssf} uses spin correlations from the central $4\times4$ unit cells (of size XC8-4).

\section{$R_1, R_2, R_3$ computed from different frequency integration window}
In the table below, we vary the energy integration window based on which we compute the ratios $R_1, R_2, R_3$ for the $S(\vec q)$ at selected momenta (Eq.3 in the main text). The ratios remain largely the same when changing the frequency integration window. The average change in each ratio is only $\sim 0.03$ for Zn-barlowite and $\sim0.01$ for herbertsmithite when enlarging the energy integration window to $\omega=[2, 6.5]$~meV. On the other hand, the boundary is where the $\delta R$ encounters an abrupt change of order one (Fig.~\ref{Fig:deltaR_line}), and is therefore little affected.

\begin{table}[h]
\begin{tabular}{l|lll|lll|}
\cline{2-7}
& Zn-barlowite                              &                                         &                      & Herbertsmithite                           &                                         &                      \\ \cline{2-7} 
& \multicolumn{1}{l|}{{[}2.5,6{]}~meV} & \multicolumn{1}{l|}{{[}2,6{]}~meV} & {[}2,6.5{]}~meV & \multicolumn{1}{l|}{{[}2.5,6{]}~meV} & \multicolumn{1}{l|}{{[}2,6{]}~meV} & {[}2,6.5{]}~meV \\ \hline
\multicolumn{1}{|l|}{$R_1$} & \multicolumn{1}{l|}{1.126}                & \multicolumn{1}{l|}{1.192}              & 1.202                & \multicolumn{1}{l|}{1.41}                 & \multicolumn{1}{l|}{1.406}              & 1.402                \\ \hline
\multicolumn{1}{|l|}{$R_2$} & \multicolumn{1}{l|}{1.430}                & \multicolumn{1}{l|}{1.410}              & 1.403                & \multicolumn{1}{l|}{1.61}                 & \multicolumn{1}{l|}{1.590}              & 1.581                \\ \hline
\multicolumn{1}{|l|}{$R_3$} & \multicolumn{1}{l|}{1.130}                & \multicolumn{1}{l|}{1.134}              & 1.128                & \multicolumn{1}{l|}{0.96}                 & \multicolumn{1}{l|}{0.955}              & 0.959                \\ \hline

\end{tabular}
\caption{$R_1, R_2, R_3$ computed from different frequency integration window}
\end{table}

\section{$J_2$ comparison}
\begin{figure}[ht]
  \includegraphics[width=0.6\linewidth]{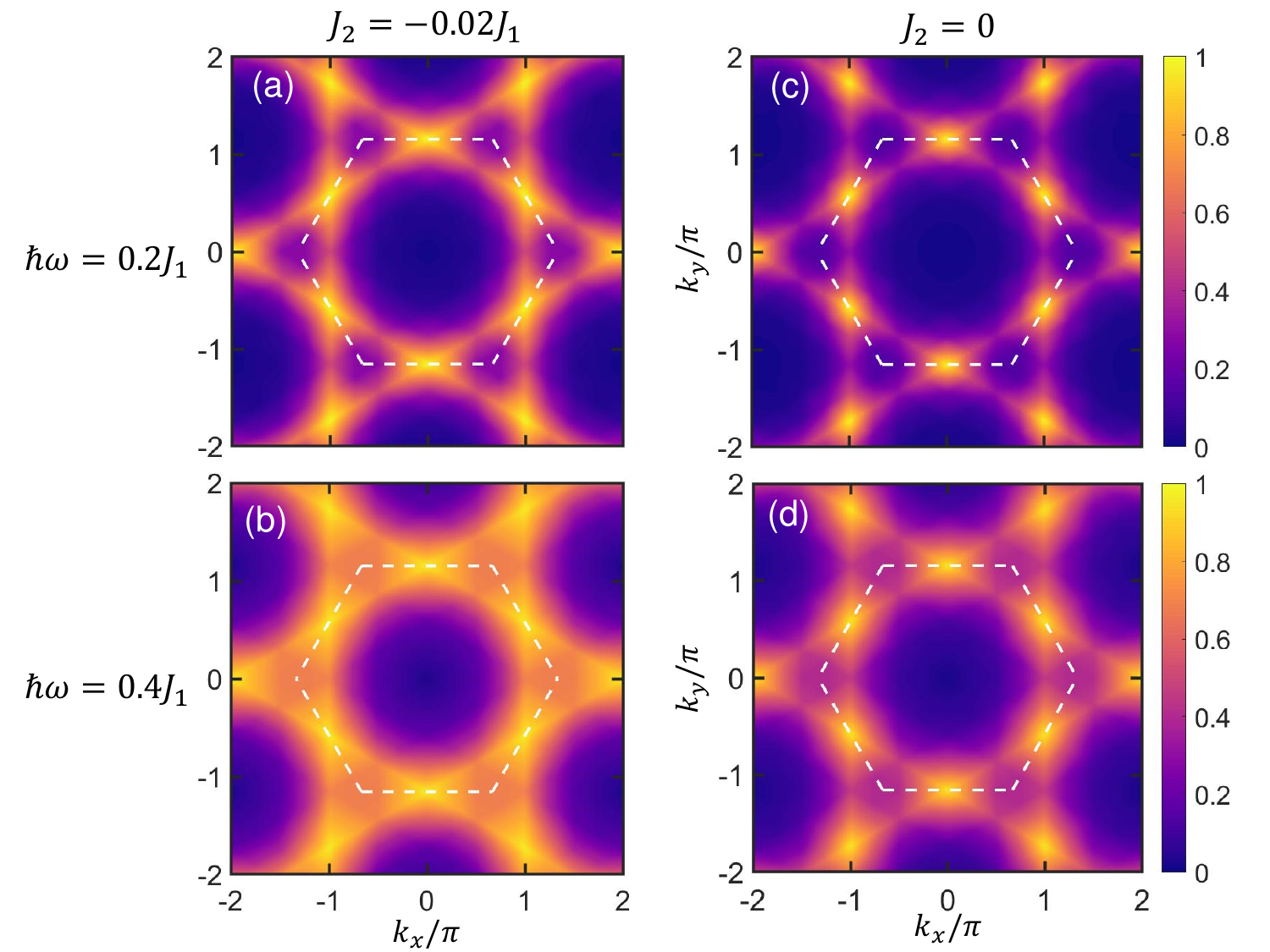}
  \caption{The $S(\mathbf{k},\omega)$ obtained by DDMRG at $\hbar\omega=0.2J_1$ (top) and $\hbar\omega=0.4J_1$ (bottom), (a)(b) with $J_2=-0.02J_1$ and (c)(d) without $J_2$. $D=0$ for all calculations. The white dashed lines are the extended BZ boundaries. The intensity is normalized to [0 1] in each figure.  The results are $D_6$ symmetrized.
  }
  \label{Fig:j2comparison}
\end{figure}

Here in Fig.~\ref{Fig:j2comparison} we compare the $S(\mathbf{k},\omega)$ obtained by DDMRG with a ferromagnetic $J_2=-0.02J_1$ and without $J_2$, both with $D=0$. It is evident that the inclusion of a ferromagnetic $J_2$ smears the peak at the $\rm \Gamma'$ point, leading to a closer resemblance with the INS data as shown in Fig.~5 in the main text.

\bibliography{Refs}